# High-resolution Tunneling Spectroscopy of ABA-stacked Trilayer Graphene

Yu Zhang, Jia-Bin Qiao, Long-Jing Yin, and Lin He[*]

Center for Advanced Quantum Studies, Department of Physics, Beijing Normal University, Beijing, 100875, People's Republic of China

*Correspondence and requests for materials should be addressed to L.H. (Email: helin@bnu.edu.cn).

**ABA-stacked trilayer graphene (TLG), the simplest system consisting of both massless and massive Dirac fermions, is expected to exhibit many interesting broken-symmetry quantum Hall states and interaction-induced phenomena. However, difficulties in microscopically identifying the stacking order of the TLG and limited spectroscopic resolution have stymied experimental probes of these interesting states and phenomena in scanning tunneling microscopy (STM) studies. Here we studied the detailed features of the electronic structure in the ABA TLG by using high-resolution STM measurements. Landau-level (LL) crossings of the massless and massive Dirac fermions, and effective mass renormalization of the massive Dirac fermions were observed, indicating strong electron-electron interactions in the ABA TLG. Most unexpectedly, we observed unconventional splittings of the lowest LLs for both the massless and massive Dirac fermions in high perpendicular magnetic fields. These splittings of the LLs, which are beyond the description of tight-binding calculations, reveal unexplored broken-symmetry quantum Hall states in the ABA TLG induced by many-body effects.**

ABA-stacked trilayer graphene (TLG), as schematically shown in Fig. 1(a), exhibits mirror reflection symmetry in structure. Its band structure, in the simplest approximation, consists of both the massless monolayer graphene (MLG)-like and massive bilayer graphene (BLG)-like bands, as shown in Fig. 1(b). In the past few years, the ABA TLG has attracted much attention because that it is an ideal platform to explore broken-symmetry quantum Hall and many-body states [1-17]. Although many interesting broken-symmetry quantum Hall states and interaction-induced phenomena have been observed in transport measurements [1-10], they are never being probed in scanning tunneling microscopy (STM) studies because of the difficulties in microscopically identifying the stacking order of the TLG and the limited spectroscopic resolution. In this work, we used the STM combined with high-magnetic field scanning tunneling spectroscopy (STS) measurement to identify the ABA TLG [18,19]. This method allows us to identify the stacking order of the TLG at nanoscale. Moreover, we carried out STM and STS measurements at temperature $T$ = 500 mK and in the presence of magnetic fields up to 15 T. The high-resolution of the spectra ($3 k_B T \approx 0.2$ meV with $k_B$ the Boltzmann constant) enables us to probe previously inaccessible interesting electronic properties of the ABA TLG in our STM studies.

In the presence of a perpendicular magnetic field $B$, the ABA TLG is expected to exhibit rich Landau level (LL) evolution [1,2,6,8-11,17-19]. The energies of LLs for massless Dirac fermions depend on the square root of the magnetic field $\sqrt{B}$, whereas for massive charge carriers they depend linearly on $B$. Therefore, the LLs from the (MLG)-like and (BLG)-like subbands in the ABA TLG should cross at some finite fields, resulting in accidental LL degeneracies at the crossing points (Fig. 1c, Fig. S2 and Fig. S3 of the Supplemental Material [20]). The unique feature of the LLs is a compelling evidence to identify the ABA TLG in experiment. In our experiment, graphene multilayers were synthesized on Ni foils by using a facile ambient pressure chemical vapor deposition (CVD) method (more details are given in Methods and in Fig. S1 of the Supplemental Material [20]). Figure 1d shows a representative STM image of the sample. In the studied region, the sample is uniform and free of defects.

Atomic-resolved STM image, as shown in inset of Fig. 1(d), shows a triangular lattice, indicating Bernal-stacked nature of the studied region, which agrees with that of the ABA TLG [18,19,21-25]. Moreover, the electronic structure of the ABA TLG in the studied region is explicitly confirmed in our STS spectra. Figure 1(e) shows the evolution of the series of Landau levels as a function of magnetic field *B* from 3 T to 15 T with a variation of 0.2 T (also see Fig. S4-S6 of the Supplemental Material [20]). Each panel of every *B* consists of 10 *dI/dV* spectra acquired at different spatial points along a line of 10 nm. The spectra recorded at different positions are reproducible (Fig. S7 of the Supplemental Material [20]), illustrating the LL evolutions are only controlled by the magnetic fields in our experiments. Such a homogenous and definite structure provides us an ideal platform to systematically study the sophisticated electronic structure of the ABA TLG. By comparing our experimental data (Fig. 1(e)) with the theoretical result (Fig. 1(c)), we can obtain all the Slonczewski–Weiss–McClure (SWMcC) parameters $\gamma_0 \sim \gamma_5$, as illustrated in Fig. 1(a). Here, the effective inversion symmetry and the degeneracy tends to be broken by next-nearest-layer coupling $\gamma_2$ and $\gamma_5$. $\gamma_3$ introduces triangular warping and $\gamma_4$ contributes to electron-hole asymmetry of the bands [16]. The obtained SWMcC parameters $\gamma_0 \sim \gamma_5$ in our experiment are listed in Table 1, which consist well with that deduced from transport measurements in previous studies. Such a result provides undoubted experimental evidence that the studied region is the ABA TLG.

According to the theoretical calculation (Fig. 1(c)), the evolutions of the LLs with magnetic field for the massless and massive Dirac fermions are mutual independence in the ABA TLG. Whereas, our experiment indicates that the LLs of the MLG-like and BLG-like subbands strongly affect each other at the LL crossings. To clear show this effect, we plotted the *dI/dV-V* spectra of the ABA TLG against the reduced energy *E/B* from 11 T to 15 T in Fig. 2(a). In the single-particle picture, all aligned peaks (blue dashed lines) are labeled with sequential LL indices of the massive Dirac fermions. However, besides the usual straight lines in the fan diagram, we observe additional non-trivial parabolic lines (red dashed lines) arising around the bias of the LL crossings. To further illustrate this phenomenon, we plotted the reduced energies of the LLs versus the magnetic fields around the crossings in Fig. 2(b) and 2(c). In contrast with the theoretical result that the reduced energies of the LLs for massive Dirac fermions should be constant, there are significant bending of the LLs near the crossings, as pointed out by arrows. Such a result, which has never been reported in previous STM experiments, is attributed to the Coulomb interactions between electrons of the LLs for massive Dirac fermions and that for massless Dirac fermions.

Our experiment demonstrated that the electron-electron interaction not only bends the LLs for massive Dirac fermions around the crossings, but also reshape the BLG-like subband of the ABA TLG. This effect can be explicitly illustrated by extracting energy dependent effective mass $m^*$ of the massive Dirac fermions. The LLs of a gapped BLG-like subband can be expressed as [26-29]:

$$E_{B,0} = E_{B,1} = E_{B,C} + \xi U/2, \qquad N = 0,1$$

$$E_{B,N} = E_{B,C} \pm \sqrt{(\hbar\omega_c)^2[n(n-1)] + (U/2)^2}, \qquad N = 2,3,4\ ...$$

where $E_{B,C}$ is the energy of BLG-like charge neutrality point (CNP), $|U| \approx E_g$, $\omega_c = eB/m^*$ is the cyclotron frequency, and $\xi = \pm 1$ are the valley indices (more details are shown in S11 of supplement materials). We can extract the effective mass $m^*$ according to both the above expressions and the measured LL sequences (Methods are shown in Fig. S11 of the Supplemental Material [20]). Figure 3(a) and 3(b) show the obtained effective mass $m^*$ versus $B$ for electrons and holes respectively. For a selected LL index, the effective mass $m^*$ on the whole increases with the magnetic fields. For a fixed magnetic field, the effective mass $m^*$ roughly increases with the LL indices. In Fig. 3(c), we summarized the effective mass $m^*$ as a function of the energy. Two main results can be obtained according to our experiment. First, with the same energy spacing away from the CNP, $m^*$ of electron is obviously larger than that of hole, yielding the ratio $m^*_{electron}/m^*_{hole} \approx 1.4$. Tight-binding calculations provide us a good description of the electron-hole asymmetry, which enables us to deduce the interlayer hopping parameter $\gamma_4 = 0.068$. Second, the $m^*$ of the massive Dirac fermions in the ABA TLG decreases with decreasing carrier densities, which demonstrates the reshaping of the BLG-like band, as schematically shown in the inset of Fig. 3(c). This phenomenon, however, fails to be described by the tight-binding Hamiltonian with the SWMcC parameters $\gamma_0 \sim \gamma_5$ in the single-particle picture, suggesting renormalization of the effective mass of the massive Dirac fermions in the ABA TLG induced by electron-electron interaction. Recently, similar Dirac cones reshaped phenomenon induced by electron-electron interaction in graphene monolayer has been demonstrated via transport experiment [30] and STM measurement [31]. Previously, transport experiment on graphene bilayer also observed effective mass renormalization of the massive Dirac fermions [32].

The high-resolution of the STS spectra allows us to detect subtle splittings of the LLs, which are closely related to the broken-symmetry quantum Hall states in the ABA TLG. In Fig. 4(a), we plot a typical STS spectrum recorded at 15 T. The LL indices of massless Dirac fermions and massive Dirac fermions are marked by the red and blue numbers respectively. A close examination of the spectrum reveals remarkable splittings of the MLG-like $N = 0$ LL and BLG-like $N = (0,1)$ LL. According to our experiment, these splittings are induced by the broken of valley degeneracies in both the MLG-like and BLG-like bands. The lifting of valley degeneracy of MLG-like $N = 0$ LL (noted by M(0−) and M(0+)) can be experimentally verified by carrying out differential conductance mapping measurements. For the $N = 0$ LL of the MLG-like subband, electronic distribution in the $K$ ($K'$) valley is equivalent to the $A(B)$ sublattice in real space (more calculations are shown in the Supplemental Material [20]) [33-35]. Insets of Fig. 4(a) show differential conductance maps recorded at 15 T at the bias voltages of the M(0−) and M(0+) LLs respectively. Both the maps exhibit triangular contrasting. The bright spots in the conductance map of the M(0−) LL correspond to the dark spots of the triangular lattice, i.e., the $A$ sites, in the STM image, whereas the bright spots in the map of the M(0+) LL correspond to the bright spots ($B$ sites) of the triangular lattice (Fig. S9 of the Supplemental Material [20] for more data). Therefore, the observed sublattice asymmetry in the maps verifies the lifting of valley degeneracy of the MLG-like $N = 0$ LL in *ABA*-stacked TLG system. For the BLG-like band, the valley-polarized B(0,1,+) LL mainly localized on the topmost graphene layer, and the other valley-polarized B(0,1,−) LL, residing on the second layer, is much weaker in the tunneling spectra because that the STM predominantly probes the top layer. Consequently, it is easy to verify that the splitting of the $N = (0,1)$ LL as the valley splitting according to their relative intensities in STS spectra.

The observed valley splitting of the lowest LL of the ABA TLG indicates that the 12-fold degeneracy (4-fold degeneracy for the $N = 0$ LL of graphene monolayer and 8-fold degeneracy for the $N = (0,1)$ LL of graphene bilayer) of the lowest LL is lifted. Such a phenomenon can be well described by the tight-binding calculation with considering finite values of $\gamma_2$ and $\gamma_5$. Unexpectedly, two important features about the splitting of the lowest LLs of both the massless and massive Dirac fermions, which are beyond the description of the tight-binding theory, are observed in our experiment. First, the energy separation between the two valley-polarized LLs, M(0−) and M(0+), is not a constant for different magnetic fields, as shown in Fig. 4(b). This contrasts to the theoretical prediction that the energy separation should be a constant and independent of magnetic field. In our experiment, the energy separation varies with the magnetic fields from 7 T to 13 T and the valley splitting of the $N = 0$ LL exhibits a dip at about 9 T and a peak at about 12 T (Fig. 4(b) and Fig. S8 of the Supplemental Material [20]). In literature, similar unconventional magnetic-field dependent valley splitting was reported previously for the $N = 1$ LL in graphene monolayer [36]. For the $N = 1$ LL, the valley splitting increases linearly with magnetic field except between 7 T and 13 T, where the valley splitting varies nonlinearly with the field and shows a dip at about 9 T and a peak at about 12 T [36]. At present, we do not have a complete explanation for the observed magnetic-field dependent valley splitting and it is not clear why both the $N = 0$ and $N = 1$ LLs show the unconventional behavior between 7 T and 13 T. Such a behavior may be caused by the enhanced electron-electron interactions and suggests existence of unexplored correlated electronic phases in graphene in the presence of large magnetic fields.

Second, our experiment demonstrated that both the B(0,1,−) LL and B(0,1,+) LL are further split into two peaks at the magnetic fields above 12 T, as shown in Fig. 4(c) and Fig. 4(d). Figure 4(e) summarizes the energy separations of the two split states within each valley as a function of magnetic fields. Here $\Delta E$ is the average value and the error bars represent the standard deviation of the values observed in our experiment. A notable linear relationship between the energy separations and magnetic fields can be clearly observed at $B > 12$ T. To quantitatively describe the linear relationship, we define the effective gyromagnetic ratio as $g^* = \mu_B^{-1}(\partial \Delta E/\partial B)$, and the linear portion yields a $g^* = 14.5$ for the B(0,1,-) LL and $g^* = 11.5$ for the B(0,1,+) LL. We can rule out Zeeman effect for electron spins as the origin of the observed splitting for the following reasons. First, the effective gyromagnetic ratio for electron spin is usually about 2. Second, the Zeeman effect should contribute equally in two valleys. Third, the Zeeman splitting should increase linearly with fields between 0 and 15 T. Although the exact nature of the states that split the LLs is unclear at present, our experiment implies emergence of new symmetry-broken quantum Hall states at $B > 12$ T in the ABA TLG.

In conclusion, we elaborately studied the detailed features of the electronic structure in the ABA TLG by using high-resolution STM measurements, which enable us to probe previously inaccessible interesting electronic properties of the ABA TLG. Both unexplored symmetry-broken quantum Hall states and interesting interaction-induced phenomena are observed in our experiment. These results provide insight into the exotic electronic behaviors in the ABA TLG.

**Acknowledgements**

This work was supported by the National Natural Science Foundation of China (Grant Nos. 11674029, 11422430, 11374035), the National Basic Research Program of China (Grants Nos. 2014CB920903, 2013CBA01603). L.H. also acknowledges support from the National Program for Support of Top-notch Young Professionals and support from "the Fundamental Research Funds for the Central Universities".

Table 1. SWMcC parameters obtained by fitting the patterns of the LLs crossings in experiment.

|  | $\gamma_0$ (eV) | $\gamma_1$ (eV) | $\gamma_2$ (eV) | $\gamma_3$ (eV) | $\gamma_4$ (eV) | $\gamma_5$ (eV) | $\delta$ (eV) |
|---|---|---|---|---|---|---|---|
| This paper | 3.1 | 0.39 | -0.028 | 0.315 | 0.068 | 0.03 | 0.02 |
| Ref. 3 | 3.1 | 0.39 | -0.028 | 0.315 | 0.041 | 0.05 | 0.046 |
| Ref. 8 | 3.1 | 0.4 | -0.02 | 0.31 | 0.0434 | 0.04 | 0.05 |
| Ref. 9 | 3.23 | 0.31 | -0.032 | 0.3 | 0.04 | 0.01 | 0.027 |
| Ref. 10 | 3.1 | 0.39 | -0.018 | 0.315 | 0.1 | 0.01 | 0.015 |
| Ref. 13 | 3.1 | 0.39 | -0.028 | 0.315 | 0.041 | 0.05 | 0.046 |
| Ref. 14 | 3.16 | 0.4 | -0.02 | 0.3 | 0.04 | 0.04 | 0.05 |
| Ref. 15 | 3.1 | 0.39 | -0.028 | 0.315 | 0.041 | 0.05 | 0.034 |

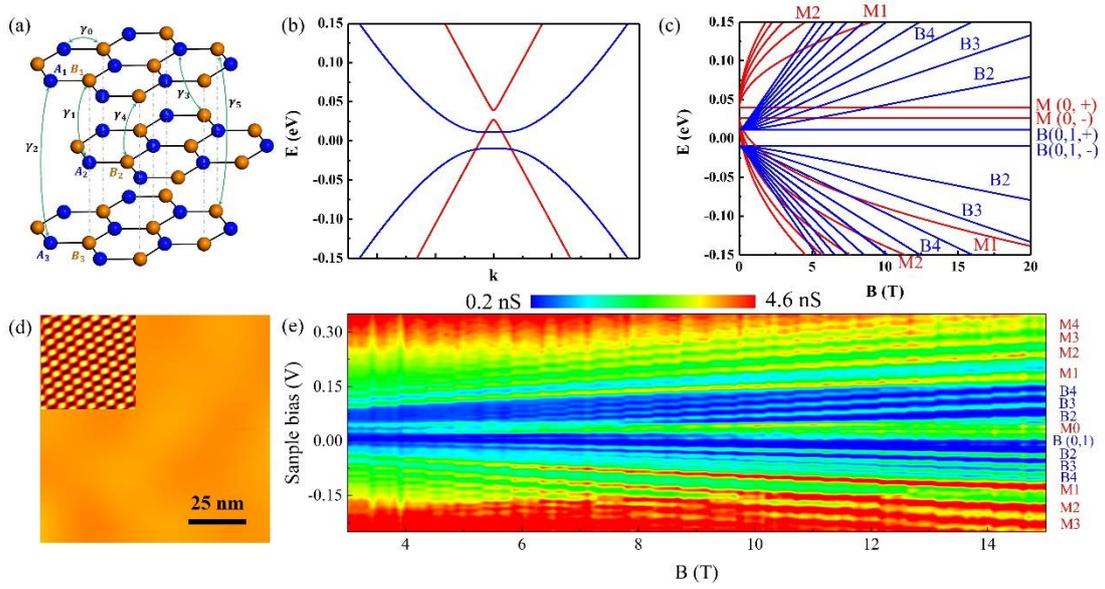

Figure 1. Electronic band structure and Landau quantization in ABA TLG. (a) Schematic diagram of ABA TLG with all the SWMcC hopping parameters. The *A* and *B* sites are denoted by blue and yellow balls, respectively. (b) Numerical calculation of a typical band structure of TLG within a full-parameter model. The red and blue lines represent the MLG-like massless Dirac fermions and BLG-like massive Dirac fermions respectively. (c) Calculations of the *B*-dependent LLs. (d) 100 nm × 100 nm STM images of decoupled ABA TLG on Ni foil. Zoom in is the atomic-resolution topography. (e) The evolution of the series of Landau levels as a function of magnetic field *B* from 3 T to 15 T as a variation of 0.2 T. Each panel of every *B* consists of 10 *dI/dV* spectra acquired at different spatial points along a line of 10 nm.

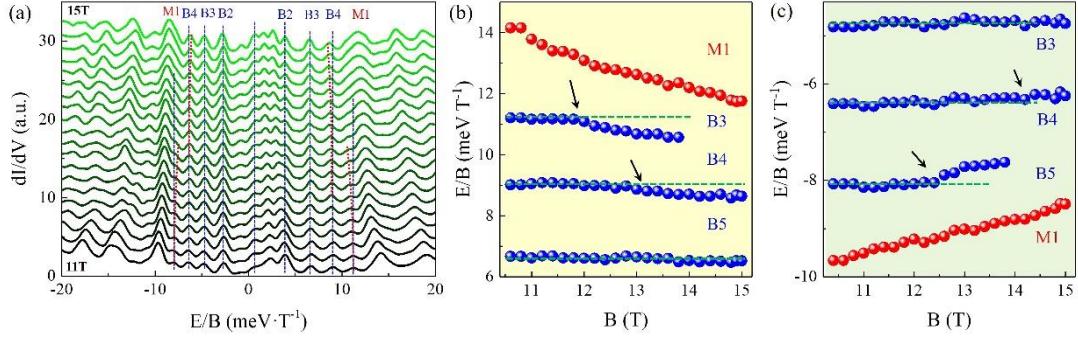

Figure 2. The strong electron-electron interactions of massless MLG-like Dirac fermions and massive BLG-like Dirac fermions through LL crossings. (a) *dI/dV-V* spectra of ABA TLG plotted against the reduced energy *E/B* from 11 T to 15 T as a variation of 0.2 T. The aligned peaks (blue dash lines) are labeled with sequential LL indices of massive Dirac fermions, and the non-trivial parabolic lines (red dash lines) arising at the bias of LL crossings. (b) LL crossing of MLG-like *N* = 1 LL and BLG-like *N* = 5 LL of holes, the significant shifts near the crossing point noted by arrow. (c) LL crossings of MLG-like *N* = 1 LL and BLG-like *N* = 5 LL of electrons in experiment.

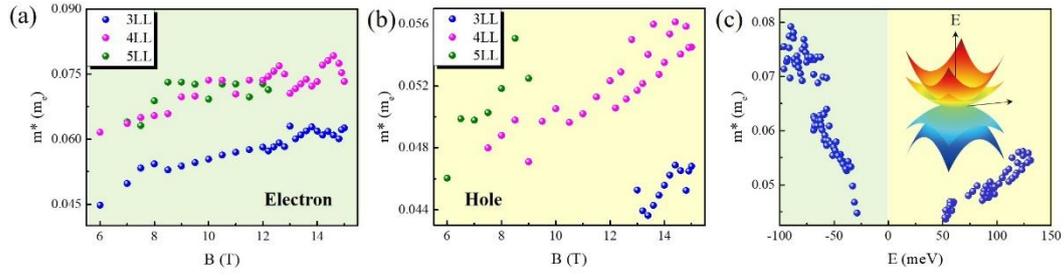

Figure 3. The electron-electron interaction in BLG-like band of ABA TLG. (a) and (b) The calculated $m^*$ versus $B$ for electrons and holes respectively. The effective mass $m^*$ of each LL are obtained from the gapped BLG-like LL sequences. (c) The relationship between $m^*$ and energy of BLG-like Dirac fermions, and the obvious electron-hole asymmetry and increasingly suppressed of $m^*$ at lower carrier densities are clearly identified. Inset: schematic diagram of BLG-like band with and without taking into account e–e interactions. The outer parabolic bands are the single-particle spectrum, and the inner parabolic bands illustrate the many-body spectrum predicted by the renormalization group theory and observed in the current experiments.

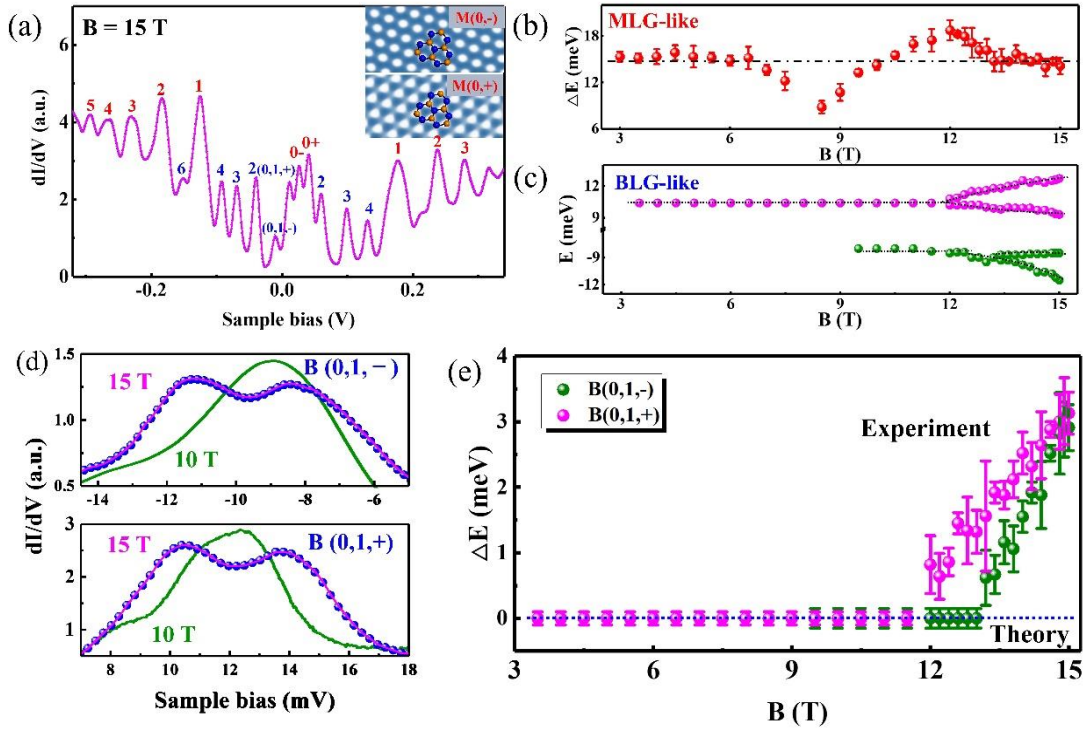

Figure 4. The broken-symmetry quantum Hall states of ABA TLG. (a) A typical STS spectra of ABA TLG recorded at 15 T. Zoom in is the differential conductance maps recorded at 15 T at the bias voltages of the M(0−) and M(0+) LLs respectively. (b) The energy separations of valleys in the MLG-like lowest LL with different perpendicular magnetic fields. (c) The B(0,1) LL evolution with magnetic fields. (d) B(0,1, −) LL and B(0,1,+) LL at 15 T and 10 T respectively, implying the symmetry-broken states at high magnetic fields. (e) The energy separations of each valley in BLG-like LL with the limit range of magnetic fields, yielding $g^*(B(0,1,-)) = 14.5$ and $g^*(B(0,1,+)) = 11.5$ for $B > 12$ T. The blue dash line shows the calculated result in Fig. 1(c).